\documentclass[journal]{IEEEtran}

\usepackage{cite}
\usepackage{amsmath}
\usepackage{amsfonts}
\usepackage{amssymb}
\usepackage{theorem}
\newtheorem{example}{Example}

\newcommand{\modulo}[2]{\left\vert #1\right\vert_{#2}}

\begin{document}

\title{Further Comments on ``Residue-to-Binary Converters Based 
       on New Chinese Remainder Theorems''}
\author{Jean-Luc~Beuchat%
\thanks{J.-L. Beuchat is with the Laboratory of Cryptography 
        and Information Security, University of Tsukuba, Japan.}%
\thanks{E-mail: beuchat@risk.tsukuba.ac.jp}}

\maketitle

\begin{abstract}
Ananda Mohan suggested that the first New Chinese Remainder Theorem
introduced by Wang can be derived from the constructive proof of the
well-known Chinese Remainder Theorem (CRT) and claimed that Wang's
approach is the same as the one proposed earlier by Huang. Ananda
Mohan's proof is however erroneous and we show here that Wang's New
CRT~I is a rewriting of an algorithm previously sketched by Hitz and
Kaltofen.
\end{abstract}

\section{Introduction}

A Residue Number System (RNS) is defined by a set of $n$ pairwise
relatively prime numbers $P_1$, \ldots, $P_n$ $\in\mathbb{N}$. Let
$M=\prod_{i=1}^nP_i$. Then, any integer $X$ belonging to
$\mathbb{Z}/M\mathbb{Z}=\{0, \ldots, M-1\}$ has a unique RNS
representation given by $(x_1 = \modulo{X}{P_1}, \ldots, x_n =
\modulo{X}{P_n})_{\text{RNS}}$, where $\modulo{X}{P_i}$ denotes
$X\text{~mod~}P_i$. The major advantage of
the RNS, which explains its popularity in digital signal processing, is
that addition, subtraction, and multiplication on large integers $X$
and $Y$ $\in\mathbb{Z}/M\mathbb{Z}$ are replaced by $n$ modular
operations performed in parallel, and whose operands are bounded by the
moduli $P_i$:
\[
\modulo{X\diamond Y}{M}=
(\modulo{x_1\diamond y_1}{P_1},\ldots,\modulo{x_n\diamond y_n}{P_n})\text{.} 
\]
where $\diamond$ denotes addition, subtraction, or
multiplication. However, an RNS is not a positional number system,
thus making conversion to integer, comparison, and division difficult
to perform. The constructive proof of the Chinese Remainder Theorem
(CRT) provides an algorithm to design a residue-to-binary converter:
\begin{equation}
X = \modulo{\sum_{i=1}^ns_i\modulo{\frac{x_i}{s_i}}{P_i}}{M}\text{,}
\label{eq:crt_1}
\end{equation}
where $s_i=M/{P_i}$ and $\modulo{1/s_i}{P_i}$ is the multiplicative
inverse of $s_i$ modulo $P_i$. The main drawback of this approach is
that it requires multiplication by the $s_i$'s, which are large
numbers, and modulo $M$ operations. The Mixed Radix System (MRS)
associated with each RNS offers another conversion scheme. A number
$X\in\mathbb{Z}/M\mathbb{Z}$ is represented by an $n$-tuple $(x'_1,
x'_2,\ldots, x'_n)_{\text{MRS}}$ such that $X=x'_1 + x_2'P_1 + x_3'P_1P_2 +
\ldots + x'_nP_1P_2\ldots P_{n-1}$, and $x'_i<P_i$, $\forall
i\in\{1,\ldots,n\}$. The Mixed Radix Conversion (MRC) is however a
strictly sequential process.

Wang~\cite{Wang:ieeetcasii-47-3} proposed a novel conversion
algorithm, called New Chinese Remainder Theorem~I (New CRT~I), based
on a new set of constants $k_i$ which are smaller than the $s_i$'s
used in the original CRT (Equation~(\ref{eq:crt_1})). He claimed that
the New CRT~I is ``substantially different from the MRC and the CRT
approaches''. A few months later, Ananda
Mohan~\cite{Mohan:ieeetcasii-47-12} suggested that the New CRT~I is
not different from the approach introduced by
Huang~\cite{Huang:ieeetc-c32-4}. His proof is unfortunately erroneous
and this paper aims at correcting it: we show that the New CRT~I is a
rewriting of an algorithm sketched by Hitz and
Kaltofen~\cite{HK:ieeetc-44-8}.

\section{Huang's Algorithm}
\label{s:huang}

Huang proposes a way to compute Equation~(\ref{eq:crt_1}) without 
modulo $M$ operations~\cite{Huang:ieeetc-c32-4}. He defines
\[
X_i = s_i\modulo{\frac{x_i}{s_i}}{P_i}\text{.}
\]
By noting that
\[
\modulo{X_i}{P_j}=
\begin{cases}
x_i & \text{if $i=j$,} \\
0   & \text{otherwise,}
\end{cases}
\]
and 
\begin{eqnarray*}
X_1 &=& (x_1, 0, 0, \ldots, 0, 0)_{\text{RNS}}\text{,} \\
X_2 &=& (0, x_2, 0, \ldots, 0, 0)_{\text{RNS}}\text{,} \\
\ldots &=& \ldots\text{,}\\
X_n &=& (0, 0, 0, \ldots, 0, x_n)_{\text{RNS}}\text{,}
\end{eqnarray*}
Huang suggests to compute the MRS representations of the $X_i$'s
by means of tables. Let $x'_{i,j}$ denote the $j$th MRS digit 
of $X_i$. It is worth noticing that $x'_{i,j}=0$ when $j<i$:
\begin{eqnarray*}
X_1 &=& (x'_{1,1}, x'_{1,2}, x'_{1,3}, \ldots, x'_{1,n-1}, x'_{1,n})_{\text{MRS}}\text{,} \\
X_2 &=& (0,        x'_{2,2}, x'_{2,3}, \ldots, x'_{2,n-1}, x'_{2,n})_{\text{MRS}}\text{,} \\
\ldots &=& \ldots\text{,}\\
X_n &=& (0, 0, 0, \ldots, 0, x'_{n,n})_{\text{MRS}}\text{.}
\end{eqnarray*}
Szabo and Tanaka's algorithm~\cite{ST:residue_arithmetic} allows one
to determine all the MRS digits of $X_i$ from $x_i$. Since
$x'_{1,1}=x_1$~\cite{ST:residue_arithmetic}, this conversion step
requires $n(n+1)/2-1$ tables. The addition of the $X_i$'s is then
performed in the MRS. Recall that the sum of two digits of weight
$P_1\ldots P_i$ generates a carry $v$ if it can be written as
$u+vP_{i+1}$, with $u< P_{i+1}$. At the end of this addition process,
Huang gets:
\[
\begin{aligned}
X &= \modulo{x_1'+x_2'P_1+\ldots+x'_nP_1\ldots P_{n-1}+qM}{M} \\
  &= x_1'+x_2'P_1+\ldots+x'_nP_1\ldots P_{n-1}\text{,}
\end{aligned}
\]  
where $q\in\mathbb{N}$ is the output carry. Since arithmetic 
is performed modulo $M$, $qM$ vanishes and Huang obtains eventually 
the MRS digits of $X$.

\begin{example}
\label{ex:conv_1}
Let $P_1=11$, $P_2=13$, and $P_3=17$. The RNS representation of
$X=514$ is then $X=(8, 7, 4)_{\text{RNS}}$. The $X_i$'s are 
defined by:
\begin{eqnarray*}
(8, 0, 0)_{\text{RNS}} &=& (8, 4, 12)_{\text{MRS}} = 1768\text{,} \\
(0, 7, 0)_{\text{RNS}} &=& (0, 3, 5)_{\text{MRS}} = 748\text{,} \\
(0, 0, 4)_{\text{RNS}} &=& (0, 0, 3)_{\text{MRS}} = 429\text{.}
\end{eqnarray*}
Let us compute the sum of the $X_i$'s in MRS. By propagating the carries,
we eventually obtain the MRS digits of $X$:
\begin{eqnarray*}
X &=& \modulo{8 + 7\cdot P_1 + 20\cdot P_1P_2}{M} \\
  &=& \modulo{8 + 7\cdot P_1 + (3+P_3)\cdot P_1P_2}{M} \\
  &=& \modulo{8 + 7\cdot P_1 + 3\cdot P_1P_2 + P_1P_2P_3}{M} \\
  &=& 8 + 7\cdot P_1 + 3\cdot P_1P_2 \\
  &=& (8, 7, 3)_{\text{MRS}} = 514\text{.}
\end{eqnarray*}
\end{example}


\section{Hitz and Kaltofen's Remark}
\label{s:hk}

The main drawback of Huang's method lies in the $n(n+1)/2-1$ tables
involved in the conversion of the $X_i$'s from RNS to MRS. Hitz and
Kaltofen~\cite{HK:ieeetc-44-8} suggest to carry out the products
$\modulo{x_i\modulo{1/s_i}{P_i}}{P_i}$ by means on $n$ modular
multipliers, where the constants $\modulo{1/s_i}{P_i}$ are
precomputed. Then, they look at the MRS representation of the
constants $s_i=(s_{i,1},\ldots,s_{i,n})_{\text{MRS}}$, and evaluate
Equation~(\ref{eq:crt_1}) in MRS. Since $s_i$ is a product of modulus,
we know that $s_{i,j}=0$ if $j<i$. Furthermore, the MRS representation
of $s_n=M/P_n$ is $(0,\ldots, 0, 1)_{\text{MRS}}$. Thus, $n(n+1)/2-1$
multiplications are required in this step. Then, it suffices to
compute the sum of the $s_i\cdot\modulo{x_i\modulo{1/s_i}{P_i}}{P_i}$
in MRS in order to get the MRS digits of $X$. Hitz and Kaltofen
describe an architecture which efficiently deals with the carries to
perform this task in~\cite{HK:ieeetc-44-8}.

\begin{example}[Example~\ref{ex:conv_1} continued]
Let us apply Hitz and Kaltofen's approach to convert $X=(8, 7, 4)_{\text{RNS}}$.
First of all, we compute the products $\modulo{x_i\modulo{1/s_i}{P_i}}{P_i}$ and 
obtain:
\[
\modulo{x_1\modulo{\frac{1}{s_1}}{P_1}}{P_1} = 8\text{,~}
\modulo{x_2\modulo{\frac{1}{s_2}}{P_2}}{P_2} = 4\text{,~}
\modulo{x_3\modulo{\frac{1}{s_3}}{P_3}}{P_3} = 3\text{.}
\]
The MRS representations of the $s_i$'s are given by:
\begin{eqnarray*}
s_1 &=& (1, 7, 1)_{\text{MRS}} = 221\text{,} \\
s_2 &=& (0, 4, 1)_{\text{MRS}} = 187\text{,} \\
s_3 &=& (0, 0, 1)_{\text{MRS}} = 143\text{.}
\end{eqnarray*}
Thus,
\begin{eqnarray*}
X &=& 8\cdot (1, 7, 1)_{\text{MRS}} +
      4\cdot (0, 4, 1)_{\text{MRS}} +
      3\cdot (0, 0, 1)_{\text{MRS}} \\
  &=& \modulo{8+72\cdot P_1 + 15\cdot P_1P_2}{M} \\
  &=& \modulo{8+(7+5\cdot P_2)\cdot P_1 + 15\cdot P_1P_2}{M} \\
  &=& \modulo{8+ 7\cdot P_1 + 20\cdot P_1P_2}{M} \\
  &=& \modulo{8+ 7\cdot P_1 + (3+P_3)\cdot P_1P_2}{M} \\
  &=& \modulo{8+ 7\cdot P_1 + 3\cdot P_1P_2 + P_1P_2P_3}{M} \\
  &=& (8, 7, 3)_{\text{MRS}} = 514\text{.}  
\end{eqnarray*}
\end{example}

Hitz and Kaltofen also point out that the multiplications by the $x_i$'s 
could be saved thanks to the second form of the CRT:
\begin{equation}
X = \modulo{\sum_{i=1}^ns_i\modulo{\frac{1}{s_i}}{P_i}x_i}{M}\text{.}
\label{eq:crt_2}
\end{equation}
Since
\[
\modulo{s_i\modulo{\frac{1}{s_i}}{P_i}}{P_j}=
\begin{cases}
1 & {\text{if $i=j$},} \\
0 & {\text{otherwise,}}
\end{cases} 
\]
we obtain another formula to compute $X$ from the $x_i$'s:
\begin{eqnarray*}
X &=& \vert x_1\cdot (1, 0, 0, \ldots, 0, 0)_{\text{RNS}} + \\
  &&  \phantom{\vert} x_2\cdot (0, 1, 0, \ldots, 0, 0)_{\text{RNS}} + \ldots + \\
  &&  \phantom{\vert} x_n\cdot (0, 0, 0, \ldots, 0, 1)_{\text{RNS}}\vert_M\text{.}
\end{eqnarray*}
The disavantage of this approach is that the
$s_i\modulo{1/s_i}{P_i}$'s are larger than the $s_i$'s, thus leading
to larger carries~\cite{HK:ieeetc-44-8}. Therefore, Hitz and Kaltofen
did not further investigate this solution. We will prove in the next
section that this algorithm turns out to be Wang's New CRT~I. It is
worth noticing that Huang stores the MRS digits of all numbers
belonging to $\mathbb{Z}/M\mathbb{Z}$, whereas Hitz and Kaltofen only
need the MRS digits of the $s_i$'s or $s_i\modulo{1/s_i}{P_i}$'s
($n(n+1)/2$ numbers).

\begin{example}[Example~\ref{ex:conv_1} continued]
Let us now convert $X=(8,7,4)_{\text{RNS}}$ according to the second form 
of the CRT given by Equation~(\ref{eq:crt_2}). Szabo and Tanaka's algorithm 
allow us to compute the MRS digits of the constants $s_i\modulo{1/s_i}{P_i}$:
\begin{eqnarray*}
s_1\modulo{1/s_1}{P_1} &=& (1, 7, 1)_{\text{MRS}} = 221\text{,} \\
s_2\modulo{1/s_2}{P_2} &=& (0, 6, 10)_{\text{MRS}} = 1496\text{,} \\
s_3\modulo{1/s_3}{P_3} &=& (0, 0, 5)_{\text{MRS}} = 715\text{.}
\end{eqnarray*}
Thus, we have
\begin{eqnarray*}
X &=& 8\cdot (1, 7, 1)_{\text{MRS}} +
      7\cdot (0, 6, 10)_{\text{MRS}} +
      4\cdot (0, 0, 5)_{\text{MRS}} \\
  &=& \modulo{8+98\cdot P_1 + 98\cdot P_1P_2}{M} \\
  &=& \modulo{8+(7+7\cdot P_2)\cdot P_1 + 98\cdot P_1P_2}{M} \\
  &=& \modulo{8+ 7\cdot P_1 + 105\cdot P_1P_2}{M} \\
  &=& \modulo{8+ 7\cdot P_1 + (3+6\cdot P_3)\cdot P_1P_2}{M} \\
  &=& \modulo{8+ 7\cdot P_1 + 3\cdot P_1P_2 + 6\cdot P_1P_2P_3}{M} \\
  &=& (8, 7, 3)_{\text{MRS}} = 514\text{.}  
\end{eqnarray*}
\end{example}

\section{Wang's New CRT~I Revisited}
\label{s:ncrti}

Wang's New CRT~I is based on the following identity~\cite{Wang:ieeetcasii-47-3}:
\begin{eqnarray}
X &=& \vert x_1 + k_1(x_2-x_1)P_1 + k_2(x_3-x_2)P_1P_2 \notag \\
  & & \phantom{\vert}+\ldots+k_{n-1}(x_n-x_{n-1})P_1\ldots P_{n-1}\vert_{M}\text{,}
\label{eq:ncrt_1}
\end{eqnarray}
where
\begin{equation}
k_i = \modulo{\frac{1}{\prod_{j=1}^iP_j}}{\prod_{j=i+1}^nP_j}\text{,}
\label{eq:def_k_i}
\end{equation}
with $1\leq i\leq n-1$. According to Wang, the New CRT~I is a fast
conversion algorithm substantially different from the CRT
approach. Let us prove that Equation~(\ref{eq:ncrt_1}) is a rewriting
of the CRT defined by Equation~(\ref{eq:crt_2}). Consider three
pairwise prime integers $a$, $b$, and $c$. The following property
holds\footnote{In order to prove this equality, it suffices to
consider the solutions of the Diophantine equation $ax + by =
\modulo{1/c}{ab}$. See for
instance~\cite{vzGG:modern_computer_algebra} for details.}:
\begin{equation}
a\modulo{\frac{1}{ac}}{b} = 
\modulo{\modulo{\frac{1}{c}}{ab}-b\modulo{\frac{1}{bc}}{a}}{ab}\text{.}
\end{equation}
If $a=P_2\ldots P_n$, $b=P_1$, and $c=1$, we have: 
\begin{eqnarray*}
s_1\modulo{\frac{1}{s_1}}{P_1} 
&=& a\modulo{\frac{1}{ac}}{b} 
= \modulo{1-P_1\modulo{\frac{1}{P_1}}{P_2\ldots P_n}}{P_1\ldots P_n} \\
&=& \modulo{1-k_1P_1}{P_1\ldots P_n}\text{.}
\end{eqnarray*}
Assume now that $2\leq i\leq n-1$, $a=P_{i+1}\ldots P_n$, $b=P_i$, and
$c=P_1\ldots P_{i=1}$. We obtain:
\begin{eqnarray*}
s_i\modulo{\frac{1}{s_i}}{P_i} 
&=& a\modulo{\frac{1}{ac}}{b}\cdot P_1\ldots P_{i-1} \\
&=& \modulo{k_{i-1}-k_iP_i}{P_i\ldots P_n}\cdot P_1\ldots P_{i-1}\text{.}
\end{eqnarray*}
Eventually, we note that:
\begin{eqnarray*}
s_n\modulo{\frac{1}{s_n}}{P_n}
&=& \modulo{\frac{1}{P_1\ldots P_{n-1}}}{P_n}\cdot P_1\ldots P_{n-1} \\
&=& k_{n-1}\cdot P_1\ldots P_{n-1}\text{.}
\end{eqnarray*}
Starting from Wang's New CRT~I, we have:
\begin{eqnarray*}
X &=& \vert x_1 + k_1(x_2-x_1)P_1 + k_2(x_3-x_2)P_1P_2  \\
  & & \phantom{\vert}+\ldots+k_{n-1}(x_n-x_{n-1})P_1\ldots P_{n-1}\vert_{M} \\
  &=& \vert (1-k_1P_1)x_1 + 
            (k_1-k_2P_2)x_2P_1 + \ldots + \\
  & & \phantom{\vert} (k_{n-2}-k_{n-1}P_{n-1})x_{n-1}P_1\ldots P_{n-2} + \\
  & & \phantom{\vert} k_{n-1}P_1\ldots P_{n-1}\vert_M \\
  &=& \vert\modulo{1-k_1P_1}{M}x_1 +
           \modulo{k_1-k_2P_2}{P_2\ldots P_n}x_2P_1 + \ldots + \\
  & & \phantom{\vert} \modulo{k_{n-2}-k_{n-1}P_{n-1}}{P_{n-1}P_n}x_{n-1}P_1\ldots P_{n-2} + \\
  & & \phantom{\vert} k_{n-1}x_nP_1\ldots P_{n-1}\vert_M \\
  &=& \modulo{\sum_{i=1}^ns_i\modulo{\frac{1}{s_i}}{P_i}x_i}{M}\text{,}
\end{eqnarray*}
which is the second form of the CRT defined by
Equation~(\ref{eq:crt_2}). Note that the $k_i$'s are the MRS digits
of numbers congruent to the $s_i\modulo{\frac{1}{s_i}}{P_i}$'s modulo
$M$. We have:
\[
\begin{aligned}
s_1\modulo{\frac{1}{s_1}}{P_1} &\equiv (1, -k_1, 0, \ldots, 0)_{\text{MRS}}\pmod{M}\text{,} \\
s_i\modulo{\frac{1}{s_i}}{P_i} &\equiv 
(0, \ldots, 0, k_{i-1}, -k_i, 0, \ldots, 0)_{\text{MRS}}\pmod{M}\text{,} \\
s_n\modulo{\frac{1}{s_n}}{P_n} &\equiv (0, \ldots, 0, k_{n-1})_{\text{MRS}}\pmod{M}\text{,}
\end{aligned}
\] 
where $2\leq i\leq n-1$. Therefore, Ananda Mohan is wrong when he
writes in~\cite{Mohan:ieeetcasii-47-12} that Wang suggests to use the
MRS representation of the $X_i$'s, which is the technique described by
Huang. We give two examples to illustrate that Wang rediscovered the
second form of the CRT (Equation~(\ref{eq:crt_2})) and that the explicit 
computation of the $k_i$'s is useless.

\begin{example}[Example~\ref{ex:conv_1} continued]
Let us convert now $X=(8,7,4)_{\text{RNS}}$ according to Equation~(\ref{eq:ncrt_1}).
We find that $k_1=201$ and $k_2=5$. We easily check that:
\[
\begin{aligned}
s_1\modulo{1/s_1}{P_1} &= (1, 7, 1)_{\text{MRS}}  &\equiv (1, -201, 0)_{\text{MRS}} \pmod{M}\text{,} \\
s_2\modulo{1/s_2}{P_2} &= (0, 6, 10)_{\text{MRS}} &\equiv (0,  201,-5)_{\text{MRS}} \pmod{M}\text{,} \\
s_3\modulo{1/s_3}{P_3} &= (0, 0, 5)_{\text{MRS}}\text{.} &
\end{aligned}
\]
In order to avoid a multiplication by $201$, Wang recommends to work with the 
following set of constants~\cite{Wang:ieeetcasii-47-3}:
\[
  a_i = 
  \begin{cases}
    \modulo{1-k_1P_1}{P_1P_2\ldots P_n}
    &\text{if $i=0$,} \\
    \modulo{k_i-k_{i+1}P_{i+1}}{P_{i+1}\ldots P_n}
    &\text{if $1\leq i\leq n-2$,} \\
    \modulo{k_{n-1}}{P_n}
    &\text{if $i=n-1$.}    
  \end{cases}
\]
He suggests to compute the MRS digits of $a_0$ and
$a_i\prod_{j=1}^iP_i$, $1\leq i\leq n-1$.  From the previous results,
it is obvious that these numbers are nothing but the
$s_i\modulo{1/s_i}{P_i}$'s. Therefore, Wang performs the following
operations:
\begin{enumerate}
\item computation of the $k_i$'s, {\sl i.e.} the MRS digits of numbers 
      congruent to the $s_i\modulo{1/s_i}{P_i}$'s modulo $M$;
\item computation of the $s_i\modulo{1/s_i}{P_i}$'s from the $k_i$'s;
\item computation of the MRS digits of the $s_i\modulo{1/s_i}{P_i}$'s according to Szabo and Tanaka's algorithm.
\end{enumerate}
Since the $s_i\modulo{1/s_i}{P_i}$'s only depend on the moduli set, it not necessary to compute the $k_i$'s.
\end{example}

\begin{example}
Wang {\em et al.}~\cite{WSAS:ieeetsp-50-7} proposed a converter for
the RNS defined by $P_1=2^n$, $P_2=2^n+1$, and $P_3=2^n-1$. They
explained that the $k_i$'s of the New CRT~I allowed them to achieve
better performance in terms of speed and area. Let us show that the
same result can be obtained without computing the $k_i$'s. Szabo and
Tanaka's conversion algorithm~\cite{ST:residue_arithmetic} provides us
with the MRS digits of the $s_i\modulo{1/s_i}{P_i}$'s:
\begin{eqnarray*}
(1, 0, 0)_{\text{RNS}} &=& (1, 1, 2^n-2)_{\text{MRS}}\text{,} \\
(0, 1, 0)_{\text{RNS}} &=& (0, 2^n, 2^{n-1}-1)_{\text{MRS}}\text{,} \\
(0, 0, 1)_{\text{RNS}} &=& (0, 0, 2^{n-1})_{\text{MRS}}\text{.}
\end{eqnarray*}
In the following, we keep the somewhat confusing notation used by 
Wang {\em et al.}~\cite{WSAS:ieeetsp-50-7}: $x_1=\modulo{X}{P_3}$,
$x_2=\modulo{X}{P_1}$, and $x_3=\modulo{X}{P_2}$. We have:
\begin{eqnarray*}
X &=& \vert x_2\cdot (1, 0, 0)_{\text{RNS}}+
            x_3\cdot (0, 1, 0)_{\text{RNS}}+ \\
  &&  \phantom{\vert} x_1\cdot (0, 0, 1)_{\text{RNS}}\vert_M \\
  &=& \vert x_2 + (x^2 + x_3\cdot 2^n)\cdot 2^n + (x_1\cdot 2^{n-1}+ \\
  && \phantom{\vert}x_2\cdot(2^n-2)+x_3\cdot(2^{n-1}-1))\cdot 2^n\cdot(2^n+1)\vert_M \\
  &=& \vert x_2 + (x^2 + x_3\cdot (2^n+1)-x_3)\cdot 2^n + (x_1\cdot 2^{n-1}+ \\
  && \phantom{\vert}x_2\cdot(-2^n)+x_3\cdot(2^{n-1}-1))\cdot 2^n\cdot(2^n+1)\vert_M \\
  &=& x_2 + 2^n\cdot\vert(x_2-x_3)+\\
  &&        (x_1-2x_2+x_3)\cdot 2^{n-1}\cdot(2^n+1)\vert_{2^{2n}-1}\text{,}
\end{eqnarray*}
which is the formula obtained by Wang {\em et al.} (see Equation~(7)
in~\cite{WSAS:ieeetsp-50-7}).
\end{example}

\section{Conclusion}

We proved that the New CRT~I was solely based on the original CRT, of
which it was only a mere rewriting, and that Wang rediscovered an
algorithm sketched by Hitz and Kaltofen in~\cite{HK:ieeetc-44-8}. We
also explained why the comment on the New CRT~I by Ananda
Mohan~\cite{Mohan:ieeetcasii-47-12} is erroneous.

\bibliographystyle{IEEEtran}
\bibliography{IEEEabrv,comments}

\end{document}